\documentclass[aps,prb,twocolumn,showpacs,groupedaddress]{revtex4}
\usepackage{amsfonts}
\usepackage{amssymb}
\usepackage{amsmath}
\usepackage{graphicx}

\usepackage{dcolumn}   
\usepackage{bm}        
\usepackage{flafter}

\begin{document}


\title{Effect of Pressure on Superconducting Ca-Intercalated Graphite CaC$_6$}

\date{\today}

\author{J. S. Kim}
\affiliation{Max-Planck-Institut f\"{u}r Festk\"{o}rperforschung,
Heisenbergstrasse 1, D-70569 Stuttgart, Germany}
\author{L. Boeri}
\affiliation{Max-Planck-Institut f\"{u}r Festk\"{o}rperforschung,
Heisenbergstrasse 1, D-70569 Stuttgart, Germany}
\author{R. K. Kremer}
\affiliation{Max-Planck-Institut f\"{u}r Festk\"{o}rperforschung,
Heisenbergstrasse 1, D-70569 Stuttgart, Germany}
\author{F. S. Razavi}
\affiliation{Department of Physics, Brock University, St.
Catharines, Ontario, L2S 3A1, Canada}

\begin{abstract}

The effect of pressure on the superconducting transition temperature
($T_c$) of the Ca-intercalated graphite compound CaC$_6$ has been
investigated up to $\sim$ 16 kbar. $T_c$ is found to increase under
pressure with a large relative ratio $\Delta$$T_c$/$T_c$ of
$\approx$ +0.4$\%$/kbar. Using first-principles calculations, we
show that the positive effect of pressure on $T_c$ can be explained
within the scope of electron-phonon theory due to the presence of a
soft phonon branch associated to in-plane vibrations of the Ca
atoms. Implications of the present findings on the current debate
about the superconducting mechanism in graphite intercalation
compounds are discussed.
\end{abstract}
\smallskip

\pacs{74.70.Ad, 74.62.Fj, 74.25.Kc, 74.62.-c}

\maketitle

\section{Introduction}

The discovery of superconductivity in Yb- and Ca-intercalated
graphites \cite{YbC6:weller:syn,CaC6:emery:syn}, with
significantly higher $T_c$'s than found previously for
alkali-metal graphite intercalation compounds (GICs) re-initiated
the debate on the long-standing puzzle of the origin of
superconductivity in GICs \cite{GIC:dresselhaus:review}. Possible
pairing mechanisms under consideration range from unconventional
exciton- or plasmon-mediated pairing \cite{GIC:csanyi:band} to the
more conventional electron-phonon (\textit{e-ph}) coupling
mechanism \cite{GIC:mazin:band,CaC6:calandra:band}.

The core of the debate on the origin of superconductivity in the
GICs are the so-called \textit{interlayer} electronic bands which
cross the Fermi level ($E_F$) besides the graphite $\pi^*$ bands.
Since the coupling of the $\pi^*$ band electrons to the graphite
phonons is not sufficient to yield the observed high $T_c$'s
\cite{C:piscanec:band,CaC6:calandra:band}, the role of the
interlayer bands and their degree of filling is considered to be
essential for superconductivity with increased $T_c$'s
\cite{GIC:csanyi:band}. Cs$\rm\acute{a}$nyi \textit{et al.}
proposed that the electrons in the interlayer bands can be
considered as nearly-free 2-dimensional (2D) electrons propagating
essentially between the graphene layers \cite{GIC:csanyi:band},
and pairing can be mediated by acoustic plasmons in these 2D
metallic slabs \cite{Plasmon:bill:theory}. However, there is a
growing body of evidence that the interlayer bands have
sufficiently strong coupling with both in-plane intercalant and
out-of-plane graphite phonon modes to allow for the relatively
high $T_c$'s within the scope of a standard \textit{e-ph} coupling
mechanism \cite{GIC:mazin:band,CaC6:calandra:band}. This scenario
gained some experimental support from recent penetration
depth\cite{CaC6:lamura:penetration}, specific
heat\cite{CaC6:jskim:Cp}, tunneling\cite{CaC6:bergeal:STM} and Ca
isotope effect\cite{CaC6:hinks:isotope} experiments in CaC$_6$.

Interestingly, the latter experiments showed a Ca isotope effect
with $\alpha_{Ca}=0.5$, suggesting a dominant role of Ca-related
modes in the superconducting pairing\cite{CaC6:hinks:isotope}.
This finding, however, is at odds with the results of
\textit{ab-initio} calculations\cite{CaC6:calandra:band} in which
isotope exponents $\alpha_{\rm Ca}$ $\approx$ $\alpha_C$ $\approx$
0.25 have been concluded. The fact that the out-of-plane phonon
modes of the host layers cannot be ignored gains substantial
support from recent investigations on other layered compounds,
\textit{e.g.} CaAlSi ($T_c$ $\sim$ 8 K), where Ca atoms are
located in between Al-Si honeycomb sheets (AlB$_2$ structure
type)\cite{CaAlSi:mazin:band}. Hence, a deeper understanding of
the superconducting pairing, and of the nature and role of the
electronic and vibrational states involved, will provide useful
insight into superconductivity in the GICs as well as other
recently discovered "high-$T_c$" superconductors.

The investigation of the pressure dependence of $T_c$ is a key
experiment to test and compare these theoretical predictions. The
degree of filling of the interlayer bands, crucial for the
acoustic plasmon pairing mechanism, depends not only on the charge
transfer from the intercalant, but also on the separation of the
graphene sheets \cite{GIC:csanyi:band}. By applying hydrostatic
pressure the graphite layer spacing, and hence the energy of the
interlayer bands, can be continuously tuned without affecting the
chemical composition. On the other hand, a comparison of the
pressure dependence of $T_c$ with the results of first-principles
calculations of the electronic and vibrational properties of the
GICs could support or rule out the hypothesis of an \textit{e-ph}
mediated coupling mechanism \cite{SC:savrasov:band}.

In this paper, we report the effect of pressure ($P \leq$ 16 kbar)
on $T_c$ of  CaC$_6$ and compare our results with
\textit{ab-initio} calculations of the electronic and vibrational
properties carried out as a function of pressure. Based on our
calculations, we argue that the increase of $T_c$ with pressure
can be understood within the scope of \textit{e-ph} coupling as a
consequence of a softening of phonon modes involving Ca in-plane
vibrations, while it is at odds with an acoustic plasmon
mechanism. We also show that there is some discrepancy between the
$T_c$ measured from experiment and that predicted by isotropic,
harmonic Migdal-Eliashberg theory, and discuss possible ways to
improve the agreement between experiments and theory.

\section{Experiment}

CaC$_6$ samples were prepared as described in detail elsewhere
\cite{CaC6:jskim:Cp,CaC6:emery:syn}. The $T_c$'s of three samples
were determined from the temperature dependence of the \textit{dc}
(samples S1 and S2) and of the \textit{ac} magnetic susceptibility
(sample S3). The variation of $T_c$ between our samples is less
than $\sim$ 0.1 K. All samples show a sharp superconducting
transition with a width $\Delta$$T_c$ $\sim$ 0.15 K between 10$\%$
and 90 $\%$ of the diamagnetic signal, indicating good sample
quality. Cu-Be piston-anvil-type pressure cells were used to apply
quasi-hydrostatic pressures with silicon oil as the pressure
transmitting medium. To monitor the pressure, $T_c$ of Pb
(99.9999\%) was determined simultaneously.

\begin{figure}[h!]
\includegraphics[width=8cm,bb=20 135 520 720]{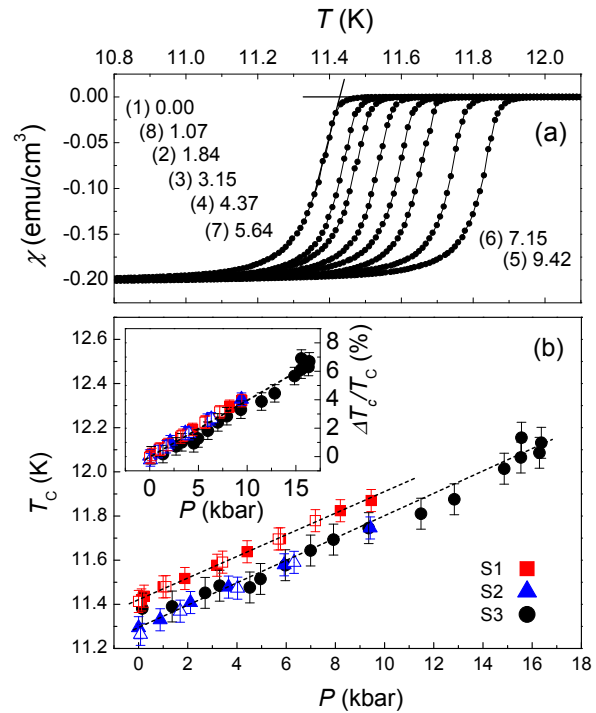}
\caption{\label{fig1}(Color online) (a) Temperature dependence of
the susceptibility for CaC$_6$ (S1) at different pressures. The
numbers next to the data and in the bracket corresponds to the
applied pressure (kbar) and the sequential order of the
measurement runs. $T_c$ is determined as the temperature where the
extrapolation of the steepest slope of $\chi(T)$ intersects the
extrapolation of the normal state $\chi(T)$ to lower temperatures.
(b) Pressure dependence of $T_c$ for three CaC$_6$ samples. The
filled and open symbols are data taken at increasing and
decreasing pressure, respectively. The dashed lines are
guide-to-eyes. The inset shows the relative change of $T_c$ with
pressure.}
\end{figure}

At ambient pressure, the onset of the superconducting transition
of our samples is $\sim$ 11.4 K, consistent with previous reports
\cite{CaC6:emery:syn}. The superconducting transition is clearly
shifted to higher temperatures as the pressure is raised (see
Fig.~\ref{fig1}(a)). Up to 16 kbar there is no indication of an
abrupt change of $T_c$ or a narrowing or broadening of the
transition, as frequently observed in other superconducting GICs,
such as \textit{e.g.} KHgC$_4$, KC$_8$, RbC$_8$, and ascribed to a
pressure-induced change of the staging or of the intercalant
sublattice \cite{GIC:clake:press}. No hysteresis in $T_c$ as well
as in the shape of $\chi(T)$ was observed between the data taken
with increasing and decreasing pressure. Thus the possible shear
stresses due to solidification of pressure medium do not
affect significantly  the superconducting properties of the
samples.

The pressure dependence of $T_c$ for three samples is shown up to
16 kbar in Fig.~\ref{fig1}(b). For all samples, $T_c$ $increases$
under pressure almost linearly with a slope of 0.042 - 0.048
K/kbar. Depending on the sample, $T_c$ and the ratio
$d$$T_c$/$d$$P$ vary slightly, but the relative change
(1/$T_c$)$d$$T_c$/$d$$P$ is the same within error bars for all
samples and amounts to $\approx$ $+$0.4 $\%$/kbar
\cite{note:YbC6:press}. Such a pressure dependence of $T_c$ is
adverse to the behavior of many $sp$ superconductors in which
often a negative pressure dependence of $T_c$ is found. Also it is
in contrast to other intercalated graphites, such as KHgC$_8$,
RbHgC$_8$ and KTl$_{1.5}$C$_4$ whose $T_c$ $decreases$ under
pressure with a rate of -2.2 $\sim$ -3.8 $\%$/kbar
\cite{GIC:clake:press}.

\section{Theory}
To gain more insight into the possible coupling mechanism we
performed a detailed Density Functional Perturbation Theory (DFPT)
\cite{DFT} study of the electronic and vibrational properties of
CaC$_6$ and explored the pressure range up to 500 kbar to
investigate whether lattice instabilities occur. In our
calculations we employed ultrasoft
pseudopotentials~\cite{Vanderbilt} with a generalized Gradient
Approximation \cite{DFT:PBE} for the exchange-correlation
functional. The eigenfunctions were expanded on a plane-wave basis
set~\cite{PWscf} with a cut-off energy of 30 Ryd and 300 Ryd for
the wavefunctions and the charge density, respectively. We used a
(8)$^3$ Monkhorst-Pack grid and a 0.06 Ryd Methfessel-Paxton
smearing, which led to convergence of better than 0.1 mRyd for the
total and 5 cm$^{-1}$ for the $\Gamma$-point frequencies. Phonon
dispersion curves were obtained by a Fourier interpolation of the
dynamical matrices calculated on a 4$^3$ grid in \textbf{q}-space
for $P$=0, 50 and 100 kbar and a 2$^3$ grid elsewhere.

The electronic structure and phonon dispersion at ambient pressure
are in excellent agreement with previous
results\cite{CaC6:calandra:band}. The Equation of State
calculations yielded equilibrium lattice constants ($a_{rh}$ =
5.16 a.u., rhombohedral angle $\theta$ = 49.90$\rm^o$) in very
good agreement with the experimental data ($a_{rh}$ = 5.17 a.u.,
$\theta$ = 49.55$\rm^o$), and a bulk modulus ($B_0 \approx$ 103
GPa) larger than that of pure graphite, but close to that of other
GICs \cite{GIC:clake:press}. The compressibility
 remains anisotropic ($k_c$/$k_a \approx$ 9) but is significantly
reduced as compared to non-intercalated graphite [see Fig. 2(b)].

\begin{figure}
\includegraphics*[width=8cm, bb=10 270 595 745]{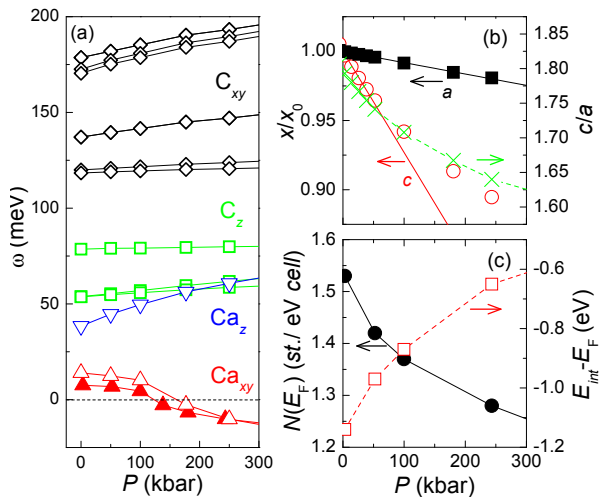}
\caption{\label{fig2}(Color online) {\em Ab-initio} pressure
dependence of selected structural, electronic and vibrational
properties of CaC$_6$. (a) Frequencies of the $\Gamma$- (open
symbols) and $X$- (solid symbols) point phonons. Imaginary
frequencies are shown as negative. The labels indicate the atom
giving the dominant contribution to the phonon eigenvector. (b)
Variation of the in-and out-of-plane lattice constants and of the
relative ratio. (c) $N(E_F)$ and position of the bottom of the
interlayer bands with respect to $E_F$.}
\end{figure}

First of all, we notice that, in general, the coupling strength
between electrons and any bosonic modes can be approximated by
$\lambda$ $\propto$
$N(E_F)$$\langle$$I^2$$\rangle$/$\langle$$\omega^2$$\rangle$,
where $N(E_F)$ is the electronic density of states at $E_F$,
$\langle$$I^2$$\rangle$ is the electron-boson matrix element, and
$\langle$$\omega^2$$\rangle$ is the square averaged frequency for
the relevant bosonic modes. Our calculations show that $N(E_F)$
\emph{decreases} with pressure because  and the interlayer band is emptied
and the $\pi^*$ bandwidth increases. (cfr.
Fig.~\ref{fig2}(c)). In the following, we therefore investigate
the alternative option to enhance $\lambda$ (and $T_c$) by a
reduction of the frequency of the relevant bosonic modes.

First, we consider electron-phonon coupling. Although unusual, the
softening of a particular phonon mode, usually an indication of an
incipient lattice instability, can induce a pressure increase in
$T_c$ \cite{note:Tc:pressure}. The calculated phonon modes for
CaC$_6$ behave differently as a function of pressure depending on
their eigenvector as seen, \textit{e.g.} in the case of the
$\Gamma$-point phonons (see Fig.~\ref{fig2}(a)): whereas the C
in-plane modes ($\omega > 100$ meV) and the Ca out-of-plane mode
($\omega \sim 30$ meV) harden with pressure, the C out-of-plane
mode ($50 < \omega < 100$ meV) are almost unaffected. On the other
hand, the lowest-lying optical mode at $\Gamma$, and an acoustical
mode at $X$ (Fig.~\ref{fig4}), both mainly involving in-plane Ca
vibrations, are considerably softened with pressure. At higher
pressures ($P$ $\gtrsim$ 120 kbar), these modes drive the system
unstable. Frozen-phonon calculations reveal that this mode couples
to both the interlayer and the $\pi^{*}$ bands, and that the
mechanism which causes its instability is similar to that giving
rise to the anharmonicity of the E$_{2g}$ mode in MgB$_2$
\cite{MgB2:boeri:band}.

\begin{figure}
\includegraphics*[width=0.99\linewidth]{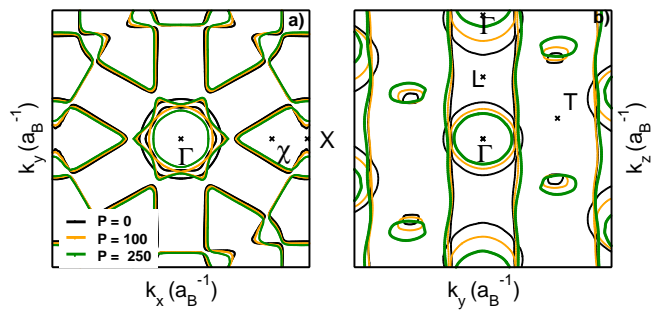}
\caption{\label{fig3}(Color online) Cross section of the Fermi
surface of CaC$_6$ at three different pressures ($P = 0, 100, 250 $
kbar), in a plane orthogonal (a) and parallel (b) to the $c$-axis.
The spherical, 3D Fermi surface centered around $\Gamma$ is due to
the interlayer band, while the 2D $\pi^*$ bands form distorted
cylinder along the $c$ axis. The position of the special {\bf k}
points in this plot are relative to the $P$ = 0 lattice constants.}
\end{figure}

To estimate the pressure dependence of the partial $\lambda_{\rm
Ca}$ associated with the Ca in-plane vibrations, we employ a
simple approximation based on the Hopfield formula: $\lambda_{i}$
$=$
$N(E_F)$$\langle$$I_{i}^2$$\rangle$/$M_i$$\langle$$\omega_{i}^2$$\rangle$,
where $\langle$$I_{i}^2$$\rangle$ is the $e$-$ph$ matrix element,
$M_i$ is the phonon mass, and $\langle$$\omega_{i}^2$$\rangle$ is
square averaged phonon frequency. Assuming that $\langle$$I_{\rm
Ca}^2$$\rangle$ is constant and $\langle$$\omega_{\rm
Ca}^2$$\rangle$ can be approximated by the square of the
lowest-lying optical phonon frequency at $\Gamma$, we find a
significant increase of $\lambda_{\rm Ca}$ by 20$\%$ at $P$ = 50
kbar, and 60$\%$ at $P$ = 100 kbar with respect to its
zero-pressure value. Since the low energy phonon modes give
dominant contribution to the coupling \cite{CaC6:calandra:band},
the increase of $\lambda_{\rm Ca}$ can be strong enough to
overcome the reduction of $N(E_F)$ and as a result, increase
$T_c$.

In contrast, it appears that the other proposed pairing mechanism,
namely acoustic plasmon-mediated pairing in the interlayer bands
\cite{GIC:csanyi:band,CaC6:smith:pressure} can hardly be
reconciled with the observed pressure dependence of $T_c$. The
basic assumption of this model is the existence of 2D metallic
layers sandwiched by dielectric layers. If this is the case, due
to the significant anisotropy of the Fermi velocity and incomplete
screening, the plasmon dispersion develops a \emph{low-energy}
acoustic branch, $\Omega_{pl}(\textbf{q})$ $\sim$ $q$, which can
provide additional pairing routes for superconductivity
\cite{Plasmon:bill:theory}.

However, the electronic band structure (\text{cfr.}
Ref.~\onlinecite{CaC6:calandra:band}) and the corresponding Fermi
surface (FS) plot (see Fig.~\ref{fig3}) reveal that the
interlayer bands are 3D in character.  Additionally, there is a
substantial charge transfer to the graphite $\pi^*$ states. Both
findings question the basic assumptions of this model already at
ambient pressure. The anisotropy of the Fermi velocity emerges
almost exclusively from the graphite $\pi^*$ bands, which display
a warped cylindrical FS parallel to the $c$ axis. With pressure,
the dispersion of the $\pi^*$ bands along the $c$ axis increases
and the 3D character of the electronic structure grows. When the
system becomes more 3D, a gap in the acoustic plasmon dispersion
is developed \cite{Plasmon:bill:theory}, and the conventional
plasmon dispersion,
$\Omega_{pl}(\textbf{q})$=$\Omega_{pl}$+$\mathcal{O}$($q^2$)
($\Omega_{pl}$ $\gg$ $k_BT_c$) is fully recovered.  Thus, even if any
coupling due to acoustic plasmons is assumed, its strength will
\emph{decrease} with pressure, adverse to our experimental
observation.

\begin{figure}
\includegraphics*[width=8.0cm, bb=25 60 590 565]{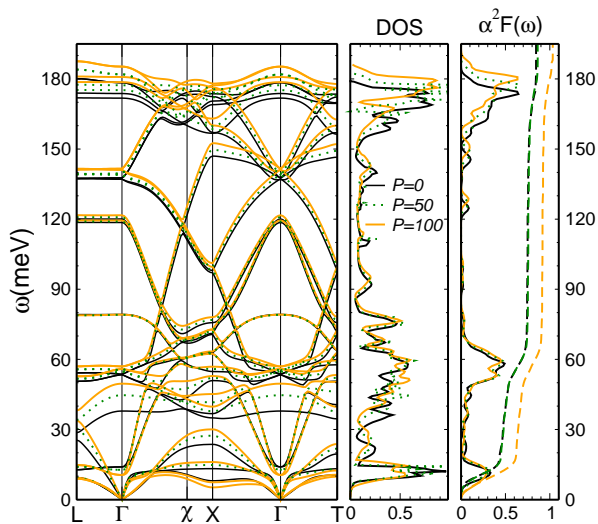}
\caption{\label{fig4}(Color online) (a) Phonon frequencies and (b)
density of states  of CaC$_6$ for $P$ = 0 (black), P=50 (green),
 $P$ = 100 kbar (orange), along selected directions
in the rhombohedral unit cell; the
line $\Gamma-X$ is contained in the graphene planes, while
$L-\Gamma$ is orthogonal to it. (c) Eliashberg function $\alpha^2
F(\Omega)$ and frequency-dependent electron-phonon coupling
$\lambda(\omega)=2\int_{0}^{\omega} \alpha^2 F(\Omega)/\Omega
d\Omega$.
 }
\end{figure}

Finally, we check whether the increase of $\lambda$ due to the
phonon softening is compatible with the observed increase of $T_c$
under pressure by calculating the $e$-$ph$ spectral function
$\alpha^2 F(\omega)$ and the frequency-dependent $e$-$ph$ coupling
parameter $\lambda(\omega) = 2 \int_{0}^{\omega} \alpha^2
F(\Omega)\Omega^{-1}d\Omega$.
 Instead of the maximum experimental
pressure ($P$ $\approx$ 16 kbar), we choose to run these
calculations at two higher pressures, $P$ = 50 and 100 kbar. We
notice, in fact, that the experimental variation of $T_c$ up to
$P$ $\approx$ 16 kbar is too small to be resolved by the accuracy
of DFT calculations. Therefore, we choose two pressures where the
calculated frequencies are all real ({\em i.e.} the system is
structurally stable), and where our approximated Hopfield equation
predicts a sizable increase in $\lambda$; the second point ($P$ =
$100$ kbar) is chosen closed to the predicted structural
transition, in order to get an estimate of the maximum attainable
$T_c$ in this system.

The $e$-$ph$ spectral function $\alpha^2 F(\omega)$ displays three
peaks well separated in energy, corresponding to different phonon
eigenvectors. The largest contribution to the total $e$-$ph$
coupling comes from the in-plane Ca phonons, which increases
$\lambda_{\rm Ca}=0.42$ to $\lambda_{\rm Ca}=0.62 (0.44)$ at $P$ =
100 (50) kbar. The second largest contribution from the C
out-of-plane vibrations remains essentially unchanged, since the
modes at $\sim$ 70 meV are insensitive to the pressure. The total
$\lambda$ increases from 0.84 at $P$ = 0 to 0.86 at $P$ = 50 kbar
and 1.03 at $P$ = 100 kbar, which results entirely from the soft
Ca in-plane modes.

We estimate the effect of pressure on $T_c$ using the McMillan
formula \cite{Tc:mcmillan:theory}:
\begin{equation}
T_c =\frac{\langle\omega_{ln}\rangle}{1.2} \exp \left[
\frac{-1.04(1+\lambda)}{\lambda-(1+0.62\lambda)\mu^*} \right],
\end{equation}
where $<$$\omega_{ln}$$>$ is the logarithmic averaged phonon
frequency and $\mu^*$ is the Coulomb pseudopotential. Inserting
the calculated $<$$\omega_{ln}$$>$, (305 K at $P$ = 0, 300 K at
$P$ = 50 kbar, and 230 K at $P$ = 100 kbar), and setting
$\mu^{*}=0.145$, which reproduces the experimental $T_c$ at zero
pressure, we estimate that $T_c$ should increase from 11.4 K to
13.5 (12) K from $P$ = 0 to $P$ = 100(50) kbar. Therefore the
effect of the phonon softening for the in-plane Ca vibration is
strong enough to overcome those of the decrease of DOS and
$<$$\omega_{ln}$$>$, and as a result, $T_c$ increases with
pressure. We also predict that the phonon softening will drive the
system unstable for $P$ $\gtrsim$ 120 kbar, thus imposing a
theoretical limit to the maximum attainable $T_c$.

\section{Discussions}

\begin{figure}
\includegraphics*[width=8.0cm, bb=50 380 520 725]{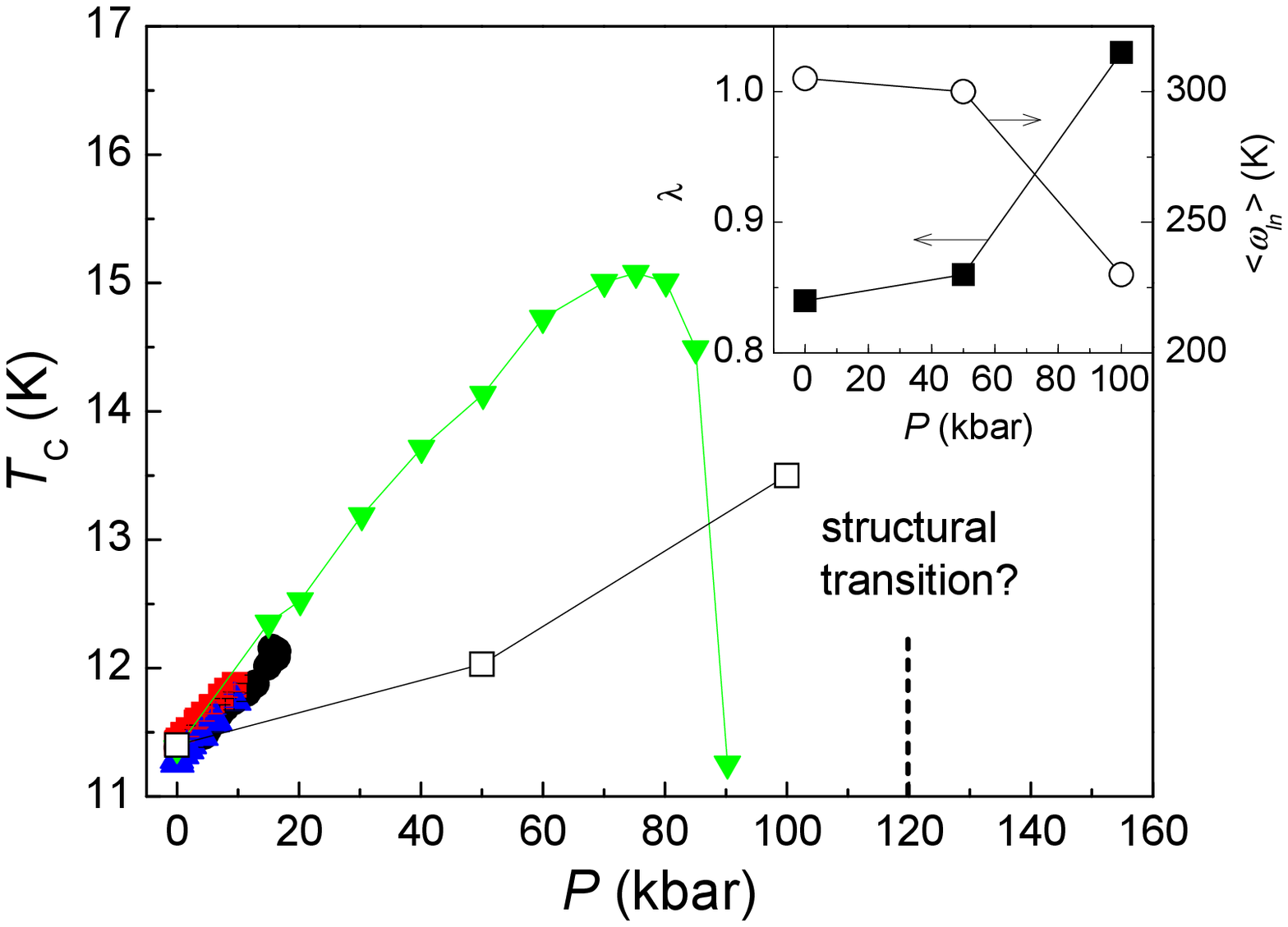}
\caption{\label{fig5}(Color online) The calculated $T_c$ (open
squares) as a function of pressure together with the measured
$T_c$ in this work (same as Fig.~\ref{fig2}) as well as Ref.
\onlinecite{CaC6:gauzzi:pressure} (green triangles). The dashed
line indicates the point at which {\em ab-initio} calculations
predict a dynamical instability of the lowest $X$-point phonon.
The inset shows a pressure dependence of the calculated $e$-$ph$
coupling constant, $\lambda$ and the logarithmic averaged phonon
frequency, $<$$\omega_{ln}$$>$. Both quantities have a strong
non-linear behavior.}
\end{figure}

After finishing the present work, an experimental study up to $P$
$\approx$ 160 kbar was posted on the cond-mat archive
\cite{CaC6:gauzzi:pressure}. Here a transition of $T_c$ at $\sim$
80 kbar was observed, accompanied by lattice softening, thus
confirming our theoretical prediction of an upper bound on $T_c$
due to lattice instability. In Fig.~\ref{fig5} we plot the
experimentally measured $T_c$, from our work and
Ref.~\onlinecite{CaC6:gauzzi:pressure}, together with the
calculated $T_c$ as a function of pressure. We notice that, even
though the positive pressure dependence is well reproduced  by our
\textit{ab-initio} calculations, there are still some
discrepancies  between experiment and theory, which we shall now
discuss in detail.

First of all, while experimentally the (structural) transition is
found at 80 kbar, we predicted it to take place at $\sim$ 120
kbar. It is well known that the structural transition pressure
for the GICs is very sensitive to  several experimental
details. For example, the previous pressure experiments on
1st-stage KC$_8$ showed large hysteresis of the transition
pressure as well as significant time dependence of the order of
days or weeks\cite{KC8:bloch:pressure}. In particular, we note
that in the GICs T$_c$ can show a strong anomaly even before
a structural transition takes place. For example, in
KC$_8$, an intercalant structural
transition occurs at $\sim$ 14 kbar\cite{KC8:fuerst:press}, but
a sudden increase of $T_c$ is observed already at $\sim$ 4 kbar
\cite{KC8:delong:press}. On the other hand, this kind of error
is also within the accuracy of DFT. While it could possibly be
reduced by a more careful convergence of the $X$ point frequency
with respect to ${\bf k}$ points sampling, there would always be
an uncertainty of 10 - 20 kbar related to the choice of the
exchange and correlation functional or of the basis
set\cite{GIC:calandra:band}. However, even if DFT can fail to
predict the exact transition pressure, it is usually much more
accurate in describing structural {\em trends} as a function of
pressure. Therefore, it is quite surprising to notice that the
behavior of $T_c$ with pressure predicted by theory deviates from
that measured experimentally: in particular, theory predicts a
much slower increase in $T_c$ ($\simeq$ 0.02 K/kbar) than
experiment, and a slightly non-linear behavior, with a stronger
increase in T$_c$ at higher pressures.

Such discrepancies indicate that the harmonic, isotropic $e$-$ph$
coupling theory may not be sufficient to describe quantitatively
the behavior of CaC$_6$ and other GICs. This confirms what was
shown by a very recent, puzzling experiments on the isotope
effect~\cite{CaC6:hinks:isotope}, which reported surprisingly high
isotope exponent for Ca, $\alpha$(Ca) $\approx$ 0.5, much higher
than the value $\alpha$(Ca) = 0.24 predicted by theory
\cite{CaC6:calandra:band}. Assuming a nontrivial contribution from
the C phonon modes, the total isotope exponent will exceed the BCS
value of 0.5. Also, discrepancies between the isotropic, harmonic
Migdal-Eliashberg theory and experiment are found in the specific
heat and the upper critical fields \cite{GICs:Mazin:review}.
Further theoretical and experimental works are highly desirable
along this direction.

The first possibility to improve the agreement between theory and
experiment would be taking into account anharmonic effects or
nonlinear coupling of the in-plane Ca phonon
modes\cite{GICs:Mazin:review}. In graphite intercalation
compounds, the vibrational modes associated to the intercalant are
extremely soft, and transitions to different in-plane sublattice
as well as staging are frequently
observed\cite{GIC:clake:press,KC8:fuerst:press}, which indicates
that these modes are strongly anharmonic.

The second approximation which may not be justified, in the
present case, is that of an isotropic electron-phonon coupling. As
we have previously discussed, the Fermi surface of CaC$_6$ is
highly anisotropic and is formed by two bands, which have very
different electronic origin ($\pi^*$ and interlayer), and are
hence coupled to different phonon modes. Therefore, different
sheets of the Fermi surface would give very different
contributions to the total electron-phonon coupling, possibly
giving rise to a "smeared" multi-gap
superconductivity\cite{CaC6:massidda:twogap}.

Finally, we would like to compare our results with a very recent
{\em ab-initio} calculations\cite{GIC:calandra:band}. Here the
authors study, using the same code\cite{PWscf}, but different
pseudopotentials, the behavior of CaC$_6$ as a function of
pressure. Similarly to us, they find a structural transition
driven from the softening of the $X$-point phonon, at some
pressure between 70 and 100 kbar, thus lower than ours and close
to the experimental value. Between 0 and 50 kbar, they find an
increase in $\lambda$ of only 0.015, and estimate an increase in
$T_c$ of $\sim$ 0.3 K; they do not calculate the $T_c$ at any
other pressure besides 0 and 50 kbar, so it is not possible to
compare directly the pressure behavior of the two sets of
calculations, but the overall qualitative picture seems to be the
same; the numerical differences in the transition pressure and
$\lambda$ are probably due to the use of different
pseudopotentials and {\bf k}-point sampling. To explain the
difference between theory and experiment, the authors of this
paper propose still another possibility, that a continuous staging
transition takes place under pressure. This is of course a
possibility which should be taken into account, but a definitive
answer in this sense could come only from X-ray or neutron
diffraction measurements under pressure.

In conclusion, we presented measurements and {\em ab-initio}
calculations of the pressure dependence of $T_c$ in CaC$_6$.
We demonstrated that the
positive pressure dependence of $T_c$ can be understood within an
$e$-$ph$ scenario due to softening of the in-plane Ca phonon
modes, while it appears not to reconcile with the acoustic plasmon
mechanism. In view of new experimental findings, we also discuss
which effects beyond the isotropic, harmonic Migdal-Eliashberg
theory would have to be taken into account to obtain a full
quantitative and qualitative agreement between theory and
experiments in CaC$_6$, and possibly other graphite intercalation
compounds.

\acknowledgments The authors acknowledge useful discussion with A.
Simon, O. K. Andersen, G. B. Bachelet, and M. Giantomassi, and
thank E. Br\"{u}cher, S. H\"{o}hn for experimental assistance.


\begin{thebibliography}{}
\bibitem{YbC6:weller:syn} T. E. Weller, M. Ellerby, S. S. Saxena, R. P. Smith, and N. T.
Skipper, Nature Physics {\bf 1}, 39 (2005).
\bibitem{CaC6:emery:syn} N. Emery, C. H$\rm \acute{e}$rold, M. d'Astuto, V. Garcia, Ch. Bellin,
J.F. Mar$\rm \hat{e}$ch$\rm \acute{e}$, P. Lagrange, and G. Loupias,
Phys. Rev. Lett. {\bf 95}, 087003 (2005).
\bibitem{GIC:dresselhaus:review} M. S. Dresselhaus and G.
Dresselhaus, Adv. Phys. {\bf 30}, 139 (1981).
\bibitem{GIC:csanyi:band} G. Cs$\rm\acute{a}$nyi, P. B. Littlewood, A. H. Nevidomskyy, C.
J. Pickard and B. D. Simons, Nature Phys. {\bf 1}, 42 (2005).
\bibitem{GIC:mazin:band} I. I. Mazin, Phys. Rev. Lett. {\bf 95}, 227001
(2005).
\bibitem{CaC6:calandra:band} M. Calandra and F. Mauri, Phys. Rev. Lett. {\bf
95}, 237002 (2005).

\bibitem{C:piscanec:band} S. Piscanec, M. Lazzeri, Francesco Mauri, A. C. Ferrari, and J. Robertson, Phys. Rev. Lett.
{\bf 93}, 185503 (2004).
\bibitem{Plasmon:bill:theory} A. Bill, H. Morawitz, V. Z. Kresin,
Phys. Rev. B {\bf68}, 144519 (2003).

\bibitem{CaC6:lamura:penetration} G. Lamura, M. Aurino, G. Cifariello, E. Di Gennaro, A. Andreone, N. Emery, C. H$\rm\acute{e}$rold, J.-F. Mar$\rm\hat{e}$ch$\rm\acute{e}$, and P. Lagrange, Phys. Rev. Lett. {\bf 96}, 107008 (2006).
\bibitem{CaC6:jskim:Cp} J. S. Kim, R. K. Kremer, L. Boeri, and F.
S. Razavi, Phys. Rev. Lett. {\bf 96}, 217002 (2006).
\bibitem{CaC6:bergeal:STM} N. Bergeal, V. Dubost, Y. Noat, W. Sacks, D.
Roditchev, N. Emery, C. H$\rm\acute{e}$rold, J.-F.
Mar$\rm\hat{e}$ch$\rm\acute{e}$, and P. Lagrange, and G. Loupias,
Phys. Rev. Lett. {\bf 97}, 077003 (2006).
\bibitem{CaC6:hinks:isotope} D.G. Hinks, D.
Rosenmann, H. Claus, M.S. Bailey, and J.D. Jorgensen,
condmat/0604642.
\bibitem{CaAlSi:mazin:band} I. I. Mazin and D. A. Papaconstnatopoulos, Phys. Rev. B {\bf 69}, 180512(R) (2004); M. Giantomassi, L. Boeri, and G. B. Bachelet, $ibid$. {\bf 72},
224512 (2005).
\bibitem{SC:savrasov:band}S. Y. Savrasov and D. Y. Savrasov
Phys. Rev. B {\bf 54}, 16487 (1996).
\bibitem{GIC:clake:press} R. Clarke and C. Uher, Adv. Phys. {\bf 33}, 469
(1984).
\bibitem{note:YbC6:press} A similar pressure dependence has also
been observed in YbC$_6$. [R.P. Smith, T.E. Weller, A.F.
Kusmartseva, N.T. Skipper, M. Ellerby, and S.S. Saxena, Physica B
{\bf378-380}, 892 (2006).]
\bibitem{DFT} W. Kohn and L. J. Sham. Phys. Rev.  {\bf 40}, A 1133,
(1965); P. Hohenberg and W. Kohn. Phys. Rev. B {\bf 136}, 864,
(1964); S. Baroni, S. de Gironcoli, A. Dal Corso, and P.
Giannozzi. Rev. Mod. Phys. {\bf73}, 515, (2001).
\bibitem{Vanderbilt} D. Vanderbilt, Phys. Rev. B {\bf 41}, R7892 (1990).
\bibitem{DFT:PBE} J. P. Perdew, K. Burke, and M. Ernzerhof, Phys. Rev. Lett. {\bf78}, 1396 (1997).
\bibitem{PWscf} S. Baroni, $et$ $al$. URL  http://www.pwscf.org.
\bibitem{note:Tc:pressure} For example, in V$_3$Si [C. W. Chu and L. R. Testardi, Phys. Rev. Lett. {\bf
32}, 766 (1974)], CaAlSi [B. Lorenz, J. Cmaidalka, R. L. Meng, and
C. W. Chu, Phys. Rev. B {\bf 68}, 014512 (2003)], and FCC Li [Deepa
Kasinathan, J. Kune$\rm\check{s}$, A. Lazicki, H. Rosner, C. S. Yoo,
R. T. Scalettar, and W. E. Pickett, Phys. Rev. Lett {\bf 96}, 047004
(2006); G. Profeta, C. Franchini, N. N. Lathiotakis, A. Floris, A.
Sanna, M. A. L. Marques, M. L$\rm\ddot{u}$ders, S. Massidda, E. K.
U. Gross, and A. Continenza, $ibid$. {\bf 96}, 047003 (2006)].
\bibitem{MgB2:boeri:band} L. Boeri, G.B. Bachelet, E. Cappelluti, and L. Pietronero
Phys. Rev. B {\bf 65}, 214501 (2002).
\bibitem{CaC6:smith:pressure} R. P. Smith, A. Kusmartseva, Y. T. C. Ko, S. S. Saxena, A. Akrap, L. Forr$\rm\acute{o}$, M. Laad, T. E. Weller, M. Ellerby, and N. T. Skipper, Phys. Rev. B {\bf74}, 024505 (2006).
\bibitem{Tc:mcmillan:theory} W. L. McMillan, Phys. Rev. {\bf 167}, 331 (1968).
\bibitem{CaC6:gauzzi:pressure} A. Gauzzi, S. Takashima, N. Takeshita, C. Terakura, H.
Takagi, N. Emery, C. H$\rm \acute{e}$rold, P. Lagrange, and G.
Loupias, cond-mat/0604204.
\bibitem{KC8:bloch:pressure} J. M. Bloch, H. Katz,
D. Moses. V. B. Cajipe. J. E. Fischer, Phys. Rev. B {\bf 31}, 6785
(1985).
\bibitem{KC8:fuerst:press} C. D. Fuerst, J. E. Fischer, J. D. Axe,
J. B. Hastings, and D. B. McWhan, Phys. Rev. Lett. {\bf 50}, 357
(1983).
\bibitem{KC8:delong:press} L. E. DeLong, V. Yeh, P. C.
Eklund, V. Tondiglia, S. E. Lambert, M. B. Maple, Phys. Rev. B
{\bf 26}, 6315 (1982).
\bibitem{GIC:calandra:band} M. Calandra and F. Mauri, cond-mat/0606372.

\bibitem{GICs:Mazin:review} I. I. Mazin, L. Boeri, O. V. Dolgov,
A. A. Golubov, G. B. Bachelet, M. Giantomassi, and O. K. Andersen,
condmat/0606404.
\bibitem{CaC6:massidda:twogap} Sandro Massidda, private communication.

\end{thebibliography}
\end{document}